\documentclass[aps,pre,reprint,superscriptaddress,amsmath,amssymb,showpacs,floatfix,longbibliography]{revtex4-1}
\usepackage{graphicx}
\usepackage{bm}
\usepackage{color}
\usepackage{lipsum}

\newcommand{\RN}[1]{\text{\uppercase\expandafter{\romannumeral#1}}}

\newcommand{\bmc}{{\bm{c}}}

\newcommand{\paths}{{\mathsf{s}[\tau]}}
\newcommand{\rpaths}{{\mathsf{s}^\dagger[\tau]}}
\newcommand{\pathc}{{\mathsf{c}[\tau]}}

\begin{document}

\title{Microscopic theory for the time irreversibility and the entropy 
production}

\author{Hyun-Myung Chun}
\affiliation{Department of Physics, University of Seoul, Seoul 02504,
Korea}
\author{Jae Dong Noh}
\affiliation{Department of Physics, University of Seoul, Seoul 02504,
Korea}
\affiliation{School of Physics, Korea Institute for Advanced Study,
Seoul 02455, Korea}

\date{\today}

\begin{abstract}
In stochastic thermodynamics, the entropy production of a thermodynamic 
system is defined by the irreversibility measured by the logarithm of 
the ratio of the path
probabilities in the forward and reverse processes.
We derive the relation between the irreversibility and the entropy
production starting from the deterministic equations of motion of the whole
system consisting of a physical system and a surrounding thermal environment. 
The physical system is driven by a nonconservative force.
The derivation assumes the Markov approximation that the environmental 
degrees of freedom equilibrate instantaneously. 
Our approach concerns the irreversibility of the whole system not only the
irreversibility of the physical system only.
This approach provides a guideline for the choice of the proper reverse process
to a given forward process.
We demonstrate our idea with an example of a charged particle in the
presence of a time-varying magnetic field.
\end{abstract}
\maketitle

\section{Introduction}
Over the past few decades, many efforts have been devoted to establishing
thermodynamics for general nonequilibrium
systems~\cite{Evans:1993tm,Gallavotti:1995wv,Jarzynski:1997uj,Oono:1998uj,
Sekimoto:1998uf,Lebowitz:1999tv,Crooks:1999ta,Hatano:2001uc,Seifert:2005vb,
Speck:2005wp}.
Among them, stochastic thermodynamics is one of the most widely used 
approaches~\cite{Seifert:2008gv,Seifert:2012es}.
In stochastic thermodynamics, dynamics of a system surrounded by a thermal
environment is described as a stochastic process governed by 
the Langevin equation or the master equation.
Thermodynamic quantities such as heat, work, and entropy production are
defined at the stochastic trajectory level in the way consistent with classical 
thermodynamics~\cite{Sekimoto:1998uf,Hatano:2001uc,Seifert:2005vb,
Esposito:2012jt}.

Suppose that a system, whose configuration is denoted by $\bm{s}$, evolves
along a stochastic path $ \paths = \{\bm{s}(t)|0\leq t\leq\tau\}$ 
in contact with a thermal environment. A time evolution is accompanied by
the entropy production, which is decomposed into the
sum $\Delta S_{\rm tot}(\paths) = \Delta S_{\rm sys}(\paths) + \Delta S_{\rm
env}(\paths)$. In stochastic thermodynamics, the system entropy change
$\Delta S_{\rm sys}$ is taken as the difference of the Shannon entropy of
the system while the environment entropy change is taken as
\begin{equation}\label{eq:defEnvEP}
\Delta S_{\text{env}}=
\ln\frac{\mathcal{P}(\paths |\bm{s}(0))}
{\mathcal{P}^\dagger(\rpaths|\bm{s}^\dagger(0))} , 
\end{equation}
where $\mathcal{P}(\paths|\bm{s}(0))$ denotes the conditional path
probability of a system following the path $\paths$ to a given initial
configuration $\bm{s}(0)$ and $\mathcal{P}^\dagger$ denotes the conditional
path probability of a system following the time reversed path $\rpaths$ to a
given initial configuration $\bm{s}^\dagger(0)$ in the reverse 
process~\cite{Lebowitz:1999tv,Seifert:2005vb,Esposito:2010bd,Lee:2013to,
Kwon:2016kc}~(detailed notations will be explained later). 
The Boltzmann constant $k_B$ is set to unity throughout the paper.
From the definition of the entropy production, stochastic thermodynamics
predicts several fluctuation theorems~\cite{Lebowitz:1999tv,Crooks:1999ta,
Hatano:2001uc,Seifert:2005vb,Speck:2005wp} for the statistical properties of
the entropy production and related quantities, which have been 
examined experimentally~\cite{Wang:2002hw,Carberry:2004fk,Trepagnier:2004wr,
Wang:2005fe, Douarche:2005tk,Douarche:2005un}.

The entropy production in \eqref{eq:defEnvEP} is written in terms of the
time irreversibility of the system. It is interesting to note that $\Delta
S_{\rm env}$ is determined by the irreversibility of the system only.
There have been several attempts to show the consistency of 
the entropy production of stochastic thermodynamics  
with that of classical thermodynamics.
The consistency was first suggested for a stochastic system by
invoking an analogy to a chemical reaction system~\cite{Schnakenberg:1976wb}.
For master equation systems, the entropy production in \eqref{eq:defEnvEP} 
is shown to be consistent with the second law of
thermodynamics~\cite{Lebowitz:1999tv}.
For Langevin equation systems, the expression in \eqref{eq:defEnvEP} 
leads to the Clausius relation $\Delta S_{\rm env} = \frac{\Delta Q}{T}$ 
where $\Delta Q$ is the heat dissipated into the thermal environment of
temperature $T$~\cite{Seifert:2005vb}.

Despite the consistency at the phenomenological level,
the entropy production in terms of the path irreversibility
still remains to be verified microscopically.
Maes and Neto\v{c}n\'{y} tried to establish the relation 
\eqref{eq:defEnvEP} for a thermal equilibrium case 
by considering Hamiltonian dynamics for a coupled
system consisting of a physical system and a surrounding 
environment~\cite{Maes:2003tc}. Under the Markov approximation 
that the degrees of freedom of the environment should equilibrate 
instantaneously, they showed that the irreversibility of the physical
system is equal to the change in the entropy of the environment.
More recently, the similar approach is applied to discrete systems
described by the master equation~\cite{Hinrichsen:2011tz,Ziener:2015kl}.

In this paper, we extend the approach of Ref.~\cite{Maes:2003tc} to
a system which is driven by an arbitrary force and surrounded 
by a thermal environment. We obtain the expression for the entropy
production starting from the deterministic equations of motion and using the
Markov approximation. The
expression is shown to be the same as the one obtained from the Langevin
equation formalism. The entropy production in
\eqref{eq:defEnvEP} depends crucially on the choice the reverse 
process. Especially, when the driving force
depends on the velocity as in the Lorentz force, different choices lead to
different expressions for the entropy production.
Our approach provides a systematic way for the proper choice of a reverse
process. We apply our approach to a charged particle in the presence of the
time-varying magnetic field.

This paper is organized as follows.
In Sec.~\ref{sec2}, we introduce the setting of the problem. We
consider deterministic Newtonian dynamics for a total that consists of 
a physical system of interest and a surrounding environment. 
The physical system is driven by a nonconservative force.
We coarse-grain the environmental degrees of freedom to derive the effective
dynamics of the system by adopting the Markov approximation.
In Sec.~\ref{sec3}, we derive the expression for the
irreversibility. We will show that the
irreversibility is the same as that obtained from the Langevin equation
approach.
In order to calculate the irreversibility, one needs to introduce a reverse 
process. We suggest a rule for the 
choice of a proper reverse process. The dependence on the choice of a reverse
process is significant when the driving force depends on the velocity. 
We explain the rule for the Lorentz force system in Sec.~\ref{sec4}.
We summarize our results in Sec.~\ref{sec:Discussion}.

\section{Coarse graining}\label{sec2}
We consider a classical system $\mathcal{S}$ 
described by $N$ Cartesian coordinates
$x_{1\leq i\leq N}$ for position and $v_{1\leq i \leq N}$ for velocity. 
The system interacts with an environment $\mathcal{E}$, which is described by
$(M-N)$ Cartesian coordinates $x_{N<i\leq M}$ and
$v_{N<i\leq M}$ for position and velocity, respectively. 
The configuration of the total system $\mathcal{U}$ 
corresponds to a point in the
$2M$-dimensional phase space $\Omega$. The phase space point is denoted by
$\bm{c} = (\bm{X},\bm{V})$ where
$\bm{X} \equiv (x_1,\cdots,x_N, x_{N+1},\cdots,x_M)$ and
$\bm{V} \equiv (v_1,\cdots,v_N,v_{N+1},\cdots,v_M)$.
Similarly, the configuration of the system $\mathcal{S}$ corresponds to
a point $\bm{s} =(\bm{x},\bm{v})$ in 
the $2N$-dimensional phase space with $\bm{x}=(x_1,\cdots,x_N)$ and
$\bm{v}=(v_1,\cdots,v_N)$.
The total system evolves in time following the deterministic
Newtonian equations of motion:
\begin{equation}\label{eq:EqofMotion}
\begin{aligned}
\dot{x}_i &=  v_i \\
\dot{v}_i &=  
   \begin{cases}
 -\frac{\partial \Phi(\bm{X})}{\partial x_i} + 
  f_i(\bm{s},\bm{\lambda}) & (1\leq i \leq N),  \vspace{2mm} \\
 -\frac{\partial \Phi(\bm{X})}{\partial x_i} & (N<i\leq M),
   \end{cases}
\end{aligned}
\end{equation}
where $\Phi(\bm{X})$ is a potential energy function of the total system 
and $\bm{f}(\bm{s},\bm{\lambda}) = 
(f_1(\bm{s},\bm{\lambda}),\cdots,f_N(\bm{s},\bm{\lambda}))$ 
is an additional nonconservative driving force applied to the system. 
It may include $L$ control parameters denoted by 
$\bm{\lambda} = \bm{\lambda}(t) = (\lambda_1(t),\cdots,\lambda_L(t))$, each
of which may depend on time.
We set all masses to be unity without loss of generality.
If the total system starts with a configuration $\bm{c}$ at time $t$, 
its subsequent state is determined uniquely by the equations of motion. Let
$\mathcal{T}_{\Delta t}(\bm{c};t)$ be the configuration after the time
interval $\Delta t$, which will be referred to as a trajectory function.

\begin{figure}
\includegraphics*[width=\columnwidth]{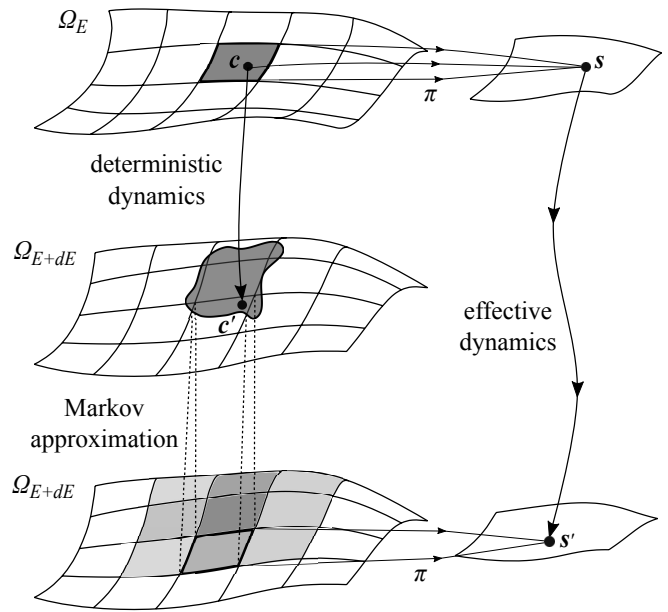}
\caption{Dynamics in the $2M$-dimensional configuration space of the 
whole system $\mathcal{U}$ and the coarse-grained $2N$-dimensional 
configuration space of the system $\mathcal{S}$. 
The constant energy surface $\Omega_E$ is divided into the subsets 
$V(\bm{s};E)$. The diagram in the left hand side represents the
deterministic time evolution of $\mathcal{U}$ followed by the equilibration 
according to the Markov approximation.
The gray scale of the shading reflects the probability density.
The darker the area is, the higher the probability density is.
}
\label{fig1}
\end{figure}

The total energy of $\mathcal{U}$ is given by $H(\bm{c}) =
\frac{1}{2}\sum_{i=1}^M v_i^2 + \Phi(\bm{X})$.
All the states of same energy $E$ constitute a constant energy
surface $\Omega_E\equiv\{\bm{c}|H(\bm{c})=E\} \subset \Omega$. 
The total energy is not conserved in the presence of the driving force. 
If $\bm{c}\in
\Omega_E$, then the configuration $\bm{c}' = \mathcal{T}_{dt}(\bm{c};t)$
belongs to another energy surface $\Omega_{E+dE}$ where 
\begin{equation}\label{delE}
dE = H(\bm{c}') - H(\bm{c}) = \sum_i f_i(\bm{s},\bm{\lambda}(t)) v_i dt .
\end{equation}
Figure~\ref{fig1} illustrates
the jump between energy surfaces.

The aim of this section is to derive the effective dynamics of the system 
out of the
deterministic dynamics of the whole system. This can be done by
coarse-graining the degrees of freedom of the environment. The most
successful method is to introduce the Markovian approximation that the degrees
of freedom of the environment equilibrate instantaneously to a given system
configuration~\cite{Maes:2003tc,Hinrichsen:2011tz}. The assumption is valid
in the limiting case where the environment relaxes
infinitely faster than the 
system~\cite{Pigolotti:2008wd,Puglisi:2010ud,Hinrichsen:2011tz,Santillan:2011gu,Esposito:2012jt,Bo:2014bq,Ziener:2015kl,Wang:2016ho}.
We adopt the Markov approximation to obtain the effective dynamics.

The coarse-graining is done by the mapping
\begin{equation}
\pi(\bm{c}) = \bm{s},
\end{equation}
which decimates the degrees of the freedom of the environment. 
For a given $\bm{c}\in \Omega_E$, the corresponding system configuration
$\bm{s} = \pi(\bm{c})$ is unique. On the other hand, there are many states in
$\Omega_E$ that are coarse-grained to the same state $\bm{s}$. The set of
all such states are denoted by
\begin{equation}
V(\bm{s};E) \equiv \{\bm{c}|\pi(\bm{c})=\bm{s} \mbox{ and } H(\bm{c})=E\} \
.
\end{equation}
These subsets are represented as the rectangular regions in
Fig.~\ref{fig1}.

We are interested in the transition probability that the system
configuration jumps from $\bm{s}$ to $\bm{s}'$ in the infinitesimal time
interval $dt$ given that the
whole system is distributed according to the probability distribution
$P(\bm{c})$ in the energy surface $\Omega_E$ initially. 
Such a transition is accompanied
with the energy change $dE = \sum_i f_i v_i dt$.
It can be written as
\begin{widetext}
\begin{equation}\label{eq:TransitionProb}
W_{dt}(\bm{s}\rightarrow \bm{s}';E,t)
= \frac{\int_{V(\bm{s};E)}d\bm{c} 
\int_{V(\bm{s}';E+dE)}d\bm{c}'
P(\bm{c}) \delta(\bm{c}'- \mathcal{T}_{dt}(\bm{c};t))}
{\int_{V(\bm{s};E)}d\bm{c} P(\bm{c})}
\end{equation}
\end{widetext}
where $\delta(\cdot)$ is the Dirac delta function, and 
$\int_{V(\bm{s};E)}d\bm{c}$ represents the 
integration over the space $V(\bm{s};E)$. 
The denominator is the probability that the
system $\mathcal{S}$ is in the configuration $\bm{s}$, while the numerator
is the joint probability that the system is at $\bm{s}$
initially and at $\bm{s}'$ after the time interval $dt$.

The Markov approximation simplifies the transition probability greatly.
Since the environment is assumed to be in the equilibrium state, $P(\bm{c})$
is uniform within each $V(\bm{s};E)$ sector~\cite{Maes:2003tc}.
Thus the factors $P(\bm{c})$ in the denominator and the numerator 
cancel each other.
The remaining factor in the numerator is equal to the volume of
$V_{dt}(\bm{s}\to\bm{s}';E,t)$ that is defined as
\begin{equation*}
V_{dt}(\bm{s}\rightarrow \bm{s}';E,t)
=\{ \bm{c} | \bm{c}\in V(\bm{s};E) \mbox{ and } 
\pi(\mathcal{T}_{dt}(\bm{c};t))=\bm{s}' \}.
\end{equation*}
It is the subset of $V(\bm{s};E)$ consisting of configurations $\bm{c} \in
V(\bm{s};E)$ that are coarse-grained to $\bm{s}'$ after
time $dt$.
Therefore, the transition probability is given by
\begin{equation}\label{eq:TransitionProb2}
W_{dt}(\bm{s}\rightarrow \bm{s}';E,t)
=\frac{|V_{dt}(\bm{s}\rightarrow \bm{s}';E,t)|}{|V(\bm{s};E)|},
\end{equation}
where $|(\cdot)|$ denotes the volume of the set $(\cdot)$ in the phase space.
The time evolution under the Markov approximation is illustrated in
Fig.~\ref{fig1}. The transition probability depends on $t$
explicitly because of the $t$ dependence of the trajectory function 
$\mathcal{T}_{dt}(\bm{c};t)$. 

\section{Irreversibility}\label{sec3}

In this section, we quantify the time irreversibility by comparing the
transition probability of a trajectory $\paths$ in a given 
dynamical process, called the forward process, with the that of 
a time-reversed trajectory denoted by 
$\rpaths = \{\bm{\epsilon s}(\tau-t) | 0 \leq t \leq \tau\}$ 
in the corresponding reverse process. 
Here, $\bm{\epsilon}$ is the time-reversal operator that
changes the sign of all the velocity coordinates. That is, $\bm{\epsilon
s} = (\bm{x},-\bm{v})$ for $\bm{s} = (\bm{x},\bm{v})$.

We first remark on the issue in defining the reverse process to a given
forward process. Consider, for example, a charged particle in the presence
of the uniform magnetic field $\bm{B}$. 
Many literatures take it granted that the
magnetic field should be flipped~($\bm{B}\to -\bm{B}$) in the reverse
process because they are the time-reversal counterpart to each 
other~\cite{Kampen:2011vs,Risken:1996vl}.
On the other hand, some studies claim that one should use the same 
field $\bm{B}$ on the ground that the irreversibility is meaningful when a
trajectory and its time-reversed trajectory are compared in the 
setting~\cite{Ganguly:2013vk,Chaudhuri:2014vf, Kwon:2016kc,Chaudhuri:2016ke}.
Such a difficulty arises when the driving force $\bm{f}$ depends explicitly 
on the velocity so that it breaks the time-reversal symmetry. 
We will provide an argument that guides us to choose the appropriate reverse
process for a general driving force $\bm{f}$. 

Consider a forward process with a driving force
$\bm{f}(\bm{s},\bm{\lambda})$ for a time interval $0\leq t \leq \tau$.
Suppose that the system evolves along a trajectory 
$\paths : \bm{s}(t_0=0) \to \cdots \to \bm{s}(t_l) \to \cdots \to
\bm{s}(t_n=\tau)$ with $t_l = l dt$.
The forward trajectory is to be compared with the time-reversed one 
$\rpaths: \bm{s}^\dagger(t_0) \to \cdots \to \bm{s}^\dagger(t_l) \to \cdots \to
\bm{s}^\dagger(t_n)$ with $\bm{s}^\dagger(t_l) = \bm{\epsilon
s}(t_{n-l}=\tau-t_l)$ in the reverse process.
Since the driving force $\bm{f}$ works on the system,
the whole system $\mathcal{U}$ jumps from one energy surface $\Omega_E$ 
to the other $\Omega_{E+dE}$ with $dE$ in \eqref{delE} in each 
step~[see also Fig.~\ref{fig1}]. 
In defining the reverse process with the
choice of the driving force $\bm{f}^\dagger(\bm{s},\bm{\lambda}^\dagger)$,
we require that not only the system $\mathcal{S}$ should return back from 
$\bm{\epsilon s}(t_{l+1})$ to $\bm{\epsilon s}(t_l)$ and but also the
whole system $\mathcal{U}$ from $\Omega_{E+dE}$ to $\Omega_{E}$ for each $l$ 
in the reverse process.
The energy surface requirement constraints the possible form of 
$\bm{f}^\dagger(\bm{s},\bm{\lambda}^\dagger)$.
The work $dE^\dagger$ done by $\bm{f}^\dagger$ in the reverse process should
cancel $dE$, which yields 
\begin{equation}
\sum_i f_i^\dagger (\bm{\epsilon s},\bm{\lambda}^{\dagger}(t)) (-dx_i) 
= -\sum_i f_i (\bm{s},\bm{\lambda}(\tau-t)) dx_i  
\end{equation}
up to the leading order in $dt$. It suggests that the driving
force in the reverse process should be chosen as
\begin{equation}\label{eq:ReverseForce2}
\bm{f}^\dagger(\bm{s},\bm{\lambda}^\dagger(t)) =
\bm{f}(\bm{\epsilon} \bm{s},\bm{\lambda}(\tau-t)).
\end{equation}

The meaning of this choice is clear.
The forces acting on the system at each time step constitute a sequence 
$\{\bm{F}_0,\ldots,\bm{F}_l,\ldots,\bm{F}_n\}$ with $\bm{F}_l =
\bm{f}(\bm{s}(t_l),\bm{\lambda}(t_l))$. 
The choice in \eqref{eq:ReverseForce2} implies that
the forces in the reverse process constitute the sequence
$\{\bm{F}^\dagger_0,\ldots,\bm{F}^\dagger_l,\ldots,\bm{F}^\dagger_n\}$ 
with $\bm{F}^\dagger_l = \bm{f}^\dagger(\bm{s}^\dagger(t_{l}),
\bm{\lambda}^\dagger(t_{l})) = \bm{f}(\bm{s}(t_{n-l}),\bm{\lambda}(t_{n-l})) =
\bm{F}_{n-l}$. 
The system is acted on by the {\em same force values} in the time-reversed 
order. 
Note that $\bm{f}^\dagger$ has a {different function form} from $\bm{f}$ 
when $\bm{f}$ depends on the velocity $\bm{v}$.  An
explicit example involving a charged particle in the presence of the
magnetic field will be discussed in Sec.~\ref{sec4}.
Another important property of the choice \eqref{eq:ReverseForce2} is that  
every trajectory $\pathc=\{\bmc (t)|0\leq t\leq\tau\}$ of the whole system 
$\mathcal{U}$ in the forward process is traced back in the reverse process. 
Formally we have
\begin{equation}\label{eq:Property}
\mathcal{T}_{t}^\dagger ( \bm{\epsilon} \mathcal{T}_{t}(\bm{c};0);\tau-t)
= \bm{\epsilon} \bm{c}
\end{equation}
with the trajectory function $\mathcal{T}^\dagger$ 
of the reverse process.

Once the reverse process is defined, the transition probability during the
infinitesimal time interval is given by
\begin{equation}
W_{dt}^\dagger(\bm{s}\to\bm{s}';E,t) = 
     \frac{ | V_{dt}^\dagger(\bm{s}\to\bm{s}';E,t) | }{ | V(s;E) | } ,
\end{equation}
where 
\begin{equation*} V_{dt}^\dagger(\bm{s}\to \bm{s}';E,t) = 
\{ \bm{c} | \bm{c}~\in V(\bm{s};E)
\mbox{ and } \pi(\mathcal{T}^\dagger_{dt}(\bm{c},t)) = \bm{s}'\} \ .
\end{equation*}
Thus, the irreversibility, given by the log ratio of the path probabilities 
as appeared in the right hand side of \eqref{eq:defEnvEP}, 
is given by the sum of 
\begin{equation}\label{eq:Irreversibility1}
dI = \ln\frac{W_{dt}(\bm{s}\to\bm{s}';E,t)}
{W_{dt}^\dagger(\bm{\epsilon}\bm{s}'\to\bm{\epsilon}\bm{s};E+dE,\tau-t)} 
= dI_1 + dI_2 \ ,
\end{equation}
where 
\begin{equation}
\begin{aligned}
dI_1 = &\ln\frac{|V(\bm{\epsilon} \bm{s}';E+dE)|}{|V(\bm{s};E)|} \\
dI_2 = &\ln\frac{|V_{dt}(\bm{s}\rightarrow \bm{s}';E,t)|}
{|V_{dt}^\dagger(\bm{\epsilon} \bm{s}'\rightarrow \bm{\epsilon} \bm{s};
E+dE,\tau-t)|}.
\end{aligned}
\end{equation}
Using the property in \eqref{eq:Property}, one finds that
$V_{dt}^\dagger(\bm{\epsilon} \bm{s}'\to\bm{\epsilon}\bm{s};E+dE,\tau-t)
=\bm{\epsilon} \mathcal{T}_{dt}(V(\bm{s}\to\bm{s}';E,t))$. 
One also finds that
$V(\bm{\epsilon s};E) = \bm{\epsilon} V(\bm{s};E)$ and that
the phase space volume is invariant under the operation of $\bm{\epsilon}$. 
Therefore, the irreversibility is given by
\begin{equation}\label{eq:Irreversibility2}
\begin{aligned}
dI_1 = & \ln\frac{|V(\bm{s}';E+dE)|}{|V(\bm{s};E)|} \\
dI_2 = & \ln\frac{|V(\bm{s}\to\bm{s}';E,t)|}
{|\mathcal{T}_{dt}(V(\bm{s}\to\bm{s}';E,t))|}.
\end{aligned}
\end{equation}
We stress that $dI$ in \eqref{eq:Irreversibility1} measures the time 
irreversibility of the whole system including the physical system and the
environment. The choice in \eqref{eq:ReverseForce2} guarantees that the
environment returns to the original energy surface in the reverse process.

The subspace $V(\bm{s};E)$ comprises the accessible states of the
environment to a given system state $\bm{s}$ in the energy surface $\Omega_E$. 
Thus, $\ln |V(\bm{s},E)|$ is the Boltzmann entropy of the
environment and $dI_1$ in Eq.~(\ref{eq:Irreversibility2}) is equal
to the change in the entropy of the environment. 
It can also be written in the Clausius form in the weak coupling limit.
The energy $E$ of the total system $\mathcal{U}$ is decomposed into 
the sum $E = E_{\rm sys} + E_{\rm env} + E_{\rm int}$, where $E_{\rm
sys}~(E_{\rm env})$ is the energy of the
system~(environment) and $E_{\rm int}$ is the interaction energy between
them. 
In the weak coupling limit, $E_{\rm int}$ is negligible so that 
$E \simeq E_{\rm sys} + E_{\rm env}$. 
Hence, we have $\ln | V(\bm{s};E) | = S_{\rm env}(E_{\rm env} = 
E-E_{\rm sys}(\bm{s}))$ and
$\ln |V(\bm{s'};E+dE)| = S_{\rm env} (E_{\rm env} = E+dE - E_{\rm sys}(\bm{s'}))$, where
$S_{\rm env}(E_{\rm env})$ denotes the entropy of the environment 
as a function of the energy. 
We note that $dE$ is the work done by the driving force on the
system. The first law of thermodynamics implies that $E_{\rm sys}(\bm{s}') -
E_{\rm sys}(\bm{s}) = dE - dQ$ where $dQ$ denotes the heat dissipated to the
environment. Consequently, we obtain that
\begin{equation}\label{dI1res}
dI_1  = \frac{dQ}{T} ,
\end{equation}
where $T = \left( \partial S_{\rm env}/\partial E_{\rm env} \right)^{-1}$ 
is the temperature of the environment. Extension to systems at strong
coupling with the environment would be interesting~\cite{Seifert:2016ik}, 
which we do not pursue in this work. 

The quantity $dI_2$ involves the expansion rate of the phase space volume
during the time evolution.  It is determined by the determinant of 
the Jacobian matrix 
$\mathsf{J} = \partial \bm{c}'/\partial \bm{c}$ with $\bm{c}' =
\mathcal{T}_{dt}(\bm{c};t)$ for $\bm{c} \in V(\bm{s};E)$.
The Jacobian matrix $\mathsf{J}$ is a block matrix of size $2M\times 2M$ in
the form of
\begin{equation}
\mathsf{J}=\begin{pmatrix}
\mathsf{A} & \mathsf{B} \\
\mathsf{C} & \mathsf{D}
\end{pmatrix}
\end{equation}
where $A_{mn}=(\partial x'_m/\partial x_n) = \delta_{mn}$,
$B_{mn}=(\partial x'_m/\partial v_n)=\delta_{mn} dt$, 
\begin{equation*}
C_{mn}=\frac{\partial v_m'}{\partial x_n} = 
\left(-\frac{\partial^2\Phi}{\partial x_m \partial x_n} 
+\sum_{i,j=1}^N\delta_{im}\delta_{jn}\frac{\partial f_i}{\partial
x_j}\right)dt ,
\end{equation*}
and 
\begin{equation}\label{eq:MatrixD}
D_{mn}=\frac{\partial v'_m}{\partial v_n} = \delta_{mn}
+\sum_{i,j=1}^N\delta_{im}\delta_{jn}\frac{\partial f_i}{\partial v_j}dt\\
\end{equation}
are the submatrices of size $M\times M$~($m,n=1,\cdots,M$) up to the first
order in $dt$, where $\delta_{mn}$ is the Kronecker delta symbol.
The determinant of the block matrix is given by 
$\det(\mathsf{J})=\det(\mathsf{D})\det(\mathsf{A}-\mathsf{B}
\mathsf{D}^{-1}\mathsf{C})$ \cite{Silvester:2000wl}.
Note that $\mathsf{A}=\mathsf{I}$, $\mathsf{B} = (dt) \mathsf{I}$, 
$\mathsf{C} = O(dt)$, and $\mathsf{D} = \mathsf{I} + \mathcal{O}(dt)$. 
Thus, we obtain that $\det(\mathsf{J})=\det(\mathsf{D}) 
=\prod_{m=1}^M D_{mm} = 1+dt\sum_{i=1}^N
\partial f_i/ \partial v_i$ up to $\mathcal{O}(dt)$, which yields that
\begin{equation}\label{dI2res}
dI_2 = \ln \det{\mathsf{J}}^{-1} = 
-dt \left(\nabla_{\bm{v}} \cdot \bm{f}\right)
\end{equation}
with the shorthand notation $(\nabla_{\bm{v}} \cdot \bm{f}) \equiv 
\sum_{i=1}^N \partial f_i / \partial v_i$.
Combining \eqref{dI1res} and \eqref{dI2res}, we finally obtain 
\begin{equation}\label{eq:Irreversibility3}
dI = \frac{dQ}{T}
-dt \left[\nabla_{\bm{v}} \cdot \bm{f}(\bm{s},\bm{\lambda})\right] \ .
\end{equation}

When the driving force does not depend on the velocity, then the
irreversibility in \eqref{eq:Irreversibility3} is equal to the change in the
entropy of the environment $dS_{\rm env}$. The same is true even in the
presence of the velocity-dependent force as long as it has the vanishing
divergence with respect to the velocity~($\nabla_{\bm{v}} \cdot \bm{f}=0$). 
The additional contribution becomes nonzero when 
$\nabla_{\bm{v}} \cdot \bm{f} \neq 0$. 
The thermodynamic meaning of the additional term remains unknown yet.

We now show that the irreversibility in \eqref{eq:Irreversibility3} based on
the deterministic dynamics incorporated with the Markovian approximation and
the weak coupling limit is reproduced in the phenomenological Langevin
equation approach.
Consider the Langevin equations
\begin{equation}\label{eq:ForwardLangevin}
\begin{aligned}
\dot{x}_i &= v_i\\
\dot{v}_i &= f_{{\rm c},i}(\bm{x}_s)
+f_i(\bm{s},\bm{\lambda})-\gamma v_i +\xi_i(t) \ .
\end{aligned}
\end{equation}
In comparison with \eqref{eq:EqofMotion}, interactions with the environment are
treated with the damping force and the thermal white noise satisfying
$\langle\xi_i(t)\rangle=0$ and
$\langle\xi_i(t)\xi_j(t')\rangle=2\gamma T \delta_{ij}\delta(t-t')$. The
system is driven by the conservative force denoted by 
$\bm{f}_{\rm c}(\bm{s})$ and the nonequililbrium driving force $\bm{f}$.
The Langevin equations for the reverse process are given by
\begin{equation}\label{eq:ReverseLangvin}
\begin{aligned}
\dot{x}_i &= v_i\\
\dot{v}_i &= f_{{\rm c},i}
+f_i^\dagger(\bm{s},\bm{\lambda}^\dagger)-\gamma v_i +\xi_i(t) \ .
\end{aligned}
\end{equation}

The Onsager-Machlup formalism allows one to write down the path probability 
for the Langevin equation system~\cite{Onsager:1953uv}. 
Using the formalism, we obtain the logarithm of the path probability ratio
of the forward and reverse processes during the infinitesimal time interval
$dt$. It is given by
\begin{equation}
\begin{aligned}\label{eq:EP_stochastic}
dI = &\frac{dQ}{T}-dt\nabla_{\bm{v}}\cdot
\left[\bm{f}(\bm{s},\bm{\lambda})-\delta{\bm{f}}\right] + 
\frac{\delta\bm{f}}{\gamma T} \circ d\bm{v} \\
& + \frac{dt}{\gamma T} \delta\bm{f} \cdot \left[-\bm{f}_{\rm c}(\bm{s}) -\bm{f}(\bm{s},\bm{\lambda})
+\delta{\bm{f}}-\gamma\bm{v}\right],
\end{aligned}
\end{equation}
where $\delta\bm{f} \equiv [\bm{f}(\bm{s},\bm{\lambda})
-\bm{f}^\dagger(\bm{\epsilon}\bm{s},
\bm{\lambda}^\dagger)]/2$ and the notation $()\circ d\bm{v}$ 
stands for the stochastic integral in the Stratonovich
sense~\cite{Gardiner:2010tp} (see Appendix~\ref{sec:appendixA} for
derivation).

When we choose the driving force $\bm{f}^\dagger$ in the reverse process
according to \eqref{eq:ReverseForce2}, 
$\delta\bm{f}$ is identically zero and the
two irreversibilities in \eqref{eq:Irreversibility3} and 
\eqref{eq:EP_stochastic} become the same. Our theory substantiates the
Langevin equation approach under the choice of \eqref{eq:ReverseForce2}.

\section{Charged particle under the Lorentz force}\label{sec4}

The irreversibility in \eqref{eq:Irreversibility1} depends crucially on the
definition of the reverse process to a given forward process. We have
proposed that the force $\bm{f}^\dagger$ should be chosen as in
\eqref{eq:ReverseForce2} on the ground that the whole system should move
back to the original energy surface in the reverse process. This choice is
characterized by the fact that the sequence of the force values 
in the reversed process is
the same as that in the forward process in the time-reversed order. 
In order to stress that the force {\em values} are the same, we refer to this
choice as the V rule.
There is an alternative choice where the
function {\em form} of the force is taken to be the
same~\cite{Ganguly:2013vk, Chaudhuri:2014vf, Kwon:2016kc, Chaudhuri:2016ke}.
It is formulated as  
\begin{equation}\label{Frule}
\bm{f}^\ddagger(\bm{s},\bm{\lambda}^\ddagger(t)) = \bm{f}(\bm{s},
\bm{\lambda}(\tau-t))
\end{equation}
In order to distinguish it from $\bm{f}^\dagger$ according to the V rule,
we use the superscript ${}^\ddagger$.
This choice will be referred to as the F rule. The merit
of the F rule is that the forward and the reverse processes are compared in
the same physical system characterized by the driving force of same form.
When the force depends on the velocity, the forces in the
reverse processes $\bm{f}^\dagger$ and $\bm{f}^\ddagger$ are different, so
are the irreversibility. In this
section, we compare the two choices for a charged particle under the Lorentz 
force.

Consider a charged particle of mass $m$ and of charge $q$ in the 
three-dimensional space with cylindrical symmetry around the $\hat{z}$
direction. The time-dependent magnetic field $\bm{B}(t) =
bt\hat{z}$ is applied to the $z$ direction with a constant $b>0$. 
According to the Maxwell equation 
$\nabla_{\bm{x}} \times \bm{E} = -\frac{\partial}{\partial t}\bm{B}$,
the time-varying magnetic field induces the electric field 
$\bm{E}(\bm{x}) = \frac{1}{2}b (y \hat{x} - x \hat{y}) = -\frac{1}{2}b
r\hat{\theta}$ with $r=\sqrt{x^2+y^2}$ and the unit vector $\hat{\theta}$
in the azimuthal direction. The electric field line circulates 
around the origin in the clockwise direction.
The particle is then applied to the Lorentz force
\begin{equation}
\bm{f}(\bm{x},\bm{v},\bm{\lambda}(t)) = 
q \bm{v} \times \bm{B}(t) + q \bm{E}(\bm{x}) \ .
\end{equation}
The field strengths are regarded as the parameters $\bm{\lambda}$.

\begin{figure}
\includegraphics*[width=\columnwidth]{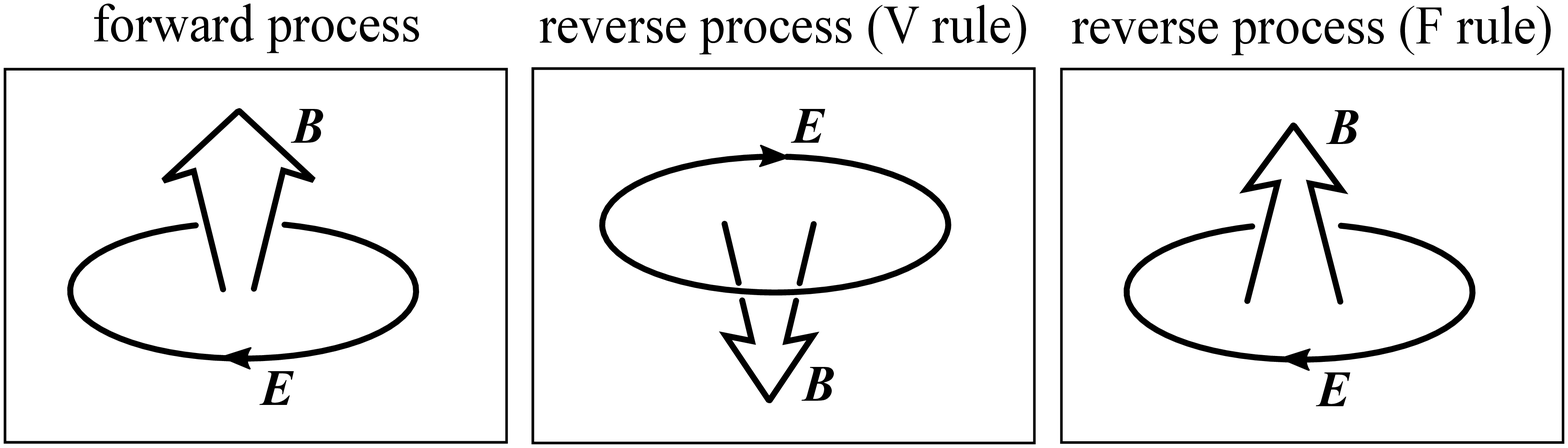}
\caption{Magnetic and electric field configurations in the forward
process~(left) and in the reverse process according to the V rule~(middle)
and the F rule~(right).
The varying width of an arrow stands for the change of the magnetic field 
strength in time.}
\label{fig2}
\end{figure}

According to the V rule the force $\bm{f}^\dagger$ in the reverse process 
is given by
\begin{equation}
\begin{aligned}
\bm{f}^\dagger(\bm{x},\bm{v},\bm{\lambda}^\dagger(t)) & = 
\bm{f}(\bm{x},-\bm{v},\bm{\lambda}(\tau-t)) \\
& =  -q \bm{v}\times \bm{B}(\tau-t) + q \bm{E}(\bm{x})  \ .
\end{aligned}
\end{equation}
It amounts to the situation that the particle is subject to the Lorentz force 
under the fields 
\begin{equation}
\bm{B}^\dagger(t) = -\bm{B}(\tau-t) , \ 
\bm{E}^\dagger(\bm{x}) = \bm{E}(\bm{x}) .
\end{equation}
Note that the magnetic field is flipped to the opposite direction.
We compare the field configurations in the forward and the reverse 
processes in Fig.~\ref{fig2}. 
The electro-magnetic fields in the reverse process
also satisfy the Maxwell's equation, 
$\nabla_{\bm x}\times \bm{E}^\dagger =
-\frac{\partial}{\partial t} \bm{B}^\dagger$.

On the other hand, the reverse process force according to the F rule,
denoted by $\bm{f}^\ddagger$, is given by
\begin{equation}
\begin{aligned}
\bm{f}^\ddagger(\bm{x},\bm{v},\bm{\lambda}^\ddagger(t)) & = 
\bm{f}(\bm{x},\bm{v},\bm{\lambda}(\tau-t)) \\
& =  q \bm{v}\times \bm{B}(\tau-t) + q \bm{E}(\bm{x})  \ .
\end{aligned}
\end{equation}
It corresponds to a Lorentz-like force under the fields
\begin{equation}
\bm{B}^\ddagger(t) = \bm{B}(\tau-t) , \ 
\bm{E}^\ddagger(\bm{x}) = \bm{E}(\bm{x}) .
\end{equation}
These fields do not satisfy the Maxwell's equation,
$\nabla_{\bm x}\times \bm{E}^\ddagger \neq
-\frac{\partial}{\partial t} \bm{B}^\ddagger$.
Namely, the reverse process in the F rule is an artificial process with
non-physical electro-magnetic fields.

The consistency with electromagnetism suggests that the V rule be
the proper way to define the reverse process for systems driven by a 
velocity-dependent force. Under the V rule, the irreversibility consists of
the Clausius entropy change of the environment and the additional term $-dt
[\nabla_{\bm{v}} \cdot \bm{f}]$. We do not know whether the additional
term can be related to any thermodynamic quantity. 
In nature, the magnetic Lorentz force is the unique example of 
a velocity-dependent force among the fundamental forces. 
If we restrict ourselves to the fundamental Lorentz force, the additional
term vanishes because the magnetic Lorentz force is divergence-free. 
Then, the irreversibility reduces to the conventional entropy production 
of the environment.
One may consider velocity-dependent forces. However, they are not the
fundamental forces but the phenomenological forces~\cite{Cerino:2015cc}.

\section{Summary}\label{sec:Discussion}
In stochastic thermodynamics, the entropy production is
given by the logarithm of the ratio of the path probabilities of the system. 
In this work, we derived
the connection between the irreversibility and the entropy production
starting from the microscopic deterministic equations of motion of the whole
system $\mathcal{U}$ consisting of a physical system $\mathcal{S}$ and an
environment $\mathcal{E}$. The key
assumption behind the connection is the Markovian approximation that the
environmental degrees of freedom equilibrates so fast that they are always
in the equilibrium state to a given configuration of $\mathcal{S}$.
Our approach is
an extension of those in Refs.~\cite{{Maes:2003tc},Hinrichsen:2011tz,
Ziener:2015kl} to systems having the continuous degrees of freedom and being
driven by an external force.
We have shown that the irreversibility derived from the microscopic point of 
view has the same expression as the entropy production of the corresponding
Langevin equation system. 

It is crucial to consider a proper reverse process to a given
forward process in characterizing the time irreversibility. In this work, we
suggest the V rule that the sequence of the force values in the reverse
process should be the same as that in the forward process in the
time-reversed order. It is formulated in \eqref{eq:ReverseForce2}. This rule
is favored because it guarantees that the whole system returns to the
original energy surface in the reverse process. This choice is contrasted to
the F rule in \eqref{Frule}, where the force in the reverse process has the
same function form as the force in the forward process. The two choices are
compared for a charged particle in the presence of time-varying magnetic
field and the induced electric field.

\begin{acknowledgments}
This work was supported by the the National Research Foundation of
Korea~(NRF) grant funded by the Korea government~(MSIP)
(No.~2016R1A2B2013972). We thank Prof. Hyunggyu Park and Prof. Chulan Kwon
for helpful discussions.
\end{acknowledgments}

\appendix
\section{Irreversibility in the Langevin system}
\label{sec:appendixA}
In this Appendix, we derive the relation \eqref{eq:EP_stochastic}
for the entropy production in the Langevin system.
The forward dynamics and the reverse dynamics of the system are governed by
Eq.~(\ref{eq:ForwardLangevin}) and Eq.~(\ref{eq:ReverseLangvin}) 
respectively.
Suppose that the system evolves from a configuration
$\bm{s}=(\bm{x},\bm{v})$ to
$\bm{s}'=(\bm{x}',\bm{v}')$ during the infinitesimal time interval $[t:t+dt]$ 
in the forward dynamics.
Such a transition occurs with the transition probability 
denoted by $W_{dt}(\bm{s}\to\bm{s}';t)$.
Similarly,
$W_{dt}^\dagger(\bm{\epsilon}\bm{s}'\to\bm{\epsilon}\bm{s};t)$
denotes the transition probability in the reverse process.
During the time interval, the control parameters change from $\bm{\lambda}(t)$ 
to $\bm{\lambda}(t+dt)$ in the forward dynamics and from 
$\bm{\lambda}^\dagger(\bar{t}-dt)$ to $\bm{\lambda}^\dagger(\bar{t})$ in the
reverse dynamics with $\bar{t}=\tau-t$.

With the help of the Onsager-Machlup formalism~\cite{Onsager:1953uv},
the transition probabilities can be written as
\begin{equation*}
\begin{aligned}
W_{dt}(\bm{s}\to\bm{s}';t)
=&\frac{\delta(d\bm{x}-\bm{v}dt)}{(4\pi\gamma T dt)^{N/2}}\\
\times &e^{-\frac{1}{4\gamma T dt}
\left\{d\bm{v} + dt\left[\nabla_{\bm{x}}\phi
-\bm{f}(\bm{s},\bm{\lambda}(t))+\gamma\bm{v}
\right]\right\}^2}
\end{aligned}
\end{equation*}
and
\begin{equation*}
\begin{aligned}
W_{dt}^\dagger(\bm{\epsilon}\bm{s}'\to\bm{\epsilon}\bm{s};\bar{t})
=&\frac{\delta(d\bm{x}-\bm{v}'dt)}{(4\pi\gamma T dt)^{N/2}}\\
\times& e^{-\frac{1}{4\gamma T dt}
\left\{d\bm{v} + dt\left[\nabla_{\bm{x}'}\phi
-\bm{f}^\dagger(\bm{\epsilon}\bm{s}',\bm{\lambda}^\dagger(\bar{t}))
-\gamma\bm{v}'\right]\right\}^2}
\end{aligned}
\end{equation*}
with $d\bm{x}=\bm{x}'-\bm{x}$ and
$d\bm{v}=\bm{v}'-\bm{v}$.
Keeping the terms up to $\mathcal{O}(dt)$, we obtain that
the irreversibility $dI = \ln W_{dt}(\bm{s}\to \bm{s}';t) /
W_{dt}^\dagger(\bm{\epsilon s}'\to \bm{\epsilon s};\bar{t})$ is given by
\begin{equation}
\begin{aligned}\label{eq:EP_app}
dI = &-\frac{\bm{v}}{T}\circ\left\{d\bm{v}
+ dt\left[\nabla_{\bm{x}}\phi-\bm{f}(\bm{s},\bm{\lambda}(t))
+\delta\bm{f}\right]\right\}\\
&+\frac{\delta\bm{f}}{\gamma T}\circ\left\{d\bm{v}
+dt\left[\nabla_{\bm{x}}\phi-\bm{f}(\bm{s},\bm{\lambda}(t))
+\delta\bm{f}\right]\right\}\\
&-dt\nabla_{\bm{v}}\cdot
\left[\bm{f}(\bm{s},\bm{\lambda}(t))-\delta\bm{f}\right]
\end{aligned}
\end{equation}
where $\delta\bm{f}
=[\bm{f}(\bm{s},\bm{\lambda}(t))-\bm{f}^\dagger
(\bm{\epsilon}\bm{s},\bm{\lambda}^\dagger(\bar{t}))]/2$ and 
the notation $d\bm{v}\circ\bm{v}
=d\bm{v}\cdot\left[\bm{v}+(\bm{v}+d\bm{v})\right]/2$ 
stands for the stochastic integral in the Stratonovich
sense~\cite{Gardiner:2010tp}. 

According to stochastic thermodynamics, the heat dissipated to the
environment is given by~\cite{Sekimoto:1998uf}
\begin{equation}\label{eq:heat_app}
dQ = \left[-d\bm{v} - dt\nabla_{\bm{x}}\phi
+dt\bm{f}(\bm{s},\bm{\lambda}(t))\right] \circ \bm{v}  .
\end{equation} 
Substituting the part in the first line in \eqref{eq:EP_app} and rearranging
all the terms, we obtain \eqref{eq:EP_stochastic}.

\bibliographystyle{apsrev}
\bibliography{paper}

\end{document}